\documentstyle[12pt,fleqn,epsfig]{article} 

\newcommand{\lsim}{\stackrel{<}{{}_\sim}}   
\newcommand{\gsim}{\stackrel{>}{{}_\sim}}

\newcommand{\lesssim}{\lsim}   
\newcommand{\be}{\begin{equation}}   
\newcommand{\ee}{\end{equation}}   
\newcommand{\bear}{\be\begin{array}}   
\newcommand{\eear}{\end{array}\ee}   
\newcommand{\bea}{\begin{eqnarray}}   
\newcommand{\eea}{\end{eqnarray}}   

\catcode`@=11   
\newtoks\@stequation   
\def\subequations{\refstepcounter{equation}%   
\edef\@savedequation{\the\c@equation}%   
  \@stequation=\expandafter{\theequation}%   %only want \theequation   
  \edef\@savedtheequation{\the\@stequation}% % expanded once   
  \edef\theequation{\theequation}%   
  \setcounter{equation}{0}%   
  \def\theequation{\theequation\alph{equation}}}   
\def\endsubequations{\setcounter{equation}{\@savedequation}%   
  \@stequation=\expandafter{\@savedtheequation}%   
  \edef\theequation{\the\@stequation}\global\@ignoretrue   
\noindent}   
\catcode`@=12   

\begin{document} 
\thispagestyle{empty} 
 
\begin{flushright}   
UCLA/99/TEP/15\\   
MPI-PhT/99-17\\      
April 1999   
\end{flushright}   
 
\begin{center}

\bigskip\bigskip 
 
{\LARGE {\bf Prompt atmospheric 
neutrinos and muons: NLO vs LO QCD predictions}} 
 
\bigskip\bigskip 
 
{\Large Graciela Gelmini\rlap{,}{$^{1}$} Paolo Gondolo\rlap{,}{$^{2}$} 
Gabriele Varieschi$^{1}$} 
 
\bigskip\bigskip 
 
${}^{1}$ {\em Dept. of Physics and Astronomy, UCLA (University of 
California, Los Angeles)\\[0pt] 
405 Hilgard Ave., Los Angeles CA 90095, USA\\[0pt] 
gelmini,variesch@physics.ucla.edu} 
 
\bigskip 
 
${}^{2}$ {\em Max-Planck-Institut f\"ur Physik (Werner-Heisenberg-Institut) 
\\[0pt] 
F\"ohringer Ring 6, 80805 M\"unchen, Germany\\[0pt] 
gondolo@mppmu.mpg.de} 
 
\bigskip\bigskip 
 
{\large {\bf Abstract}} 
\end{center} 
 
{\narrower We compare the leading and next-to-leading order QCD predictions 
for  the flux of atmospheric muons and neutrinos from decays of charmed 
particles.  We find that the full NLO lepton fluxes can be approximated to 
within $\sim  10\%$ by the Born--level fluxes multiplied by an overall 
factor of $2.2-2.4$,  which depends slightly on the PDF. This supports the 
approach in Thunman,  Ingelman, Gondolo (1996). We also find that their very 
low lepton fluxes are  due to the mild slope they used for the gluon 
distribution function at small  momentum fractions, and that substantially 
larger lepton fluxes result when  the slope of the gluon distribution 
function at small momentum fractions is  larger. } 
 
\newpage 
 
%%%%%%%%%%%%%%%%%%%%%%%%%%%%%  TEXT  %%%%%%%%%%%%%%%%%%%%%%%%%%%%%%   
%%%%%%%%%%%%%%%%%%%%%%%%%%%%%%%%%%%%%%%%%%%%%%%%%%%%%%%%%%%%%%%%%%%   
 
%\setcounter{page}{1} %\baselineskip=2\baselineskip   
 
\section{Introduction} 
 
\label{sect:intro} 
 
The flux of atmospheric neutrinos and muons at very high energies, above 1 
TeV, passes from being originated in the decays of pions and kaons to being 
predominantly generated in semileptonic decays of charmed particles (see for 
example \cite{g1}). This flux is of importance for large area detectors of 
high energy cosmic neutrinos. Future ${\rm {km}^{3}}$ arrays would be able 
to observe muons and neutrinos with energies that may reach $10^{12}$ GeV. 
Atmospheric muons and neutrinos would be one of the most important 
backgrounds, limiting the sensitivity of any ``neutrino telescope'' to 
astrophysical signals. Besides, they might be used for detector calibration 
and perhaps, more interestingly, be exploited to do physics, e.g. study 
neutrino masses. 
 
Present experimental attempts to detect atmospheric muons from charm are 
spoiled by systematic errors. Theoretical predictions depend strongly on the 
reliability of the model adopted for charm production and decay and differ 
by orders of magnitude, due to the necessity of extrapolating present 
accelerator data on open charm production in fixed target experiments, at 
laboratory energies of about 200 GeV, to the larger energies needed for 
atmospheric neutrinos, from 10$^{3}$ to ${10^{8}}$ GeV (at about $10^{8}$ 
GeV the rates become too small for a ${\rm {km^{3}}}$ detector). These 
energies, from 40 GeV to 14 TeV in the center of mass, are comparable to the 
energies of the future RHIC at Brookhaven, 200 GeV, and LHC at CERN, 7 TeV. 
 
The theoretically preferred model, perturbative QCD (pQCD), was thought to 
be inadequate because it could not account for several aspects of some of 
the early data on open charm production (in conflict with each other, on the 
other hand \cite{kvd}), and because of a sensitivity of the leading-order 
(LO) calculation, the only existing until recently, to the charm quark mass, 
to the low partonic momentum fraction, $x$, behavior of the parton 
distributions and to higher order corrections. So, even if some now-obsolete 
pQCD calculations have appeared \cite{ik,zhv}, the models for charm 
production traditionally favored in studies of atmospheric fluxes have been 
non-perturbative: for example, besides semi-empirical parametrizations of 
the cross section, the quark-gluon string model (QGSM, a.k.a. dual parton 
model), based on Regge asymptotics, and the recombination quark-parton model 
(RQPM), incorporating the assumption of an intrinsic charm component in the 
nucleon (see \cite{br}). 
 
Today, however, pQCD predictions and experimental data are known to be 
compatible \cite{a, alves1, alves2, ada, fmnr}: charm production experiments 
form a consistent set of data, and the inclusion of next-to-leading order 
(NLO) terms has been a major improvement over the leading-order treatment. 
Quoting from Appel \cite{a}, ``the success of these calculations has removed 
the impetus to look for unconventional sources of charm production beyond 
the basic QCD". 
 
A study based on pQCD was therefore performed in Ref. \cite{TIG} (called TIG 
from now on). CLEO and HERA results were incorporated, but for simplicity 
the LO charm production cross section was adopted, multiplied by a constant $ 
K$ factor of 2 to bring it in line with the next-to-leading order values, 
and supplemented by parton shower evolution and hadronization according to 
the Lund model. The neutrino and muon fluxes from charm were found to be 
lower than the lowest previous prediction, namely a factor of 20 below the 
RQPM \cite{bnsz}, of 5 below the QGSM \cite{c, gghv}, and of 3 below the 
lowest curve in Ref. \cite{zhv}. 
 
Here we use the same treatment of TIG, except for the very important 
difference of using the actual next-to-leading order pQCD calculations of 
Mangano, Nason and Ridolfi \cite{MNR} (called MNR from now on), as contained 
in the program we obtained from them (see also \cite{nde}), to compute the 
charm production cross sections. These are the same calculations used 
currently to compare pQCD predictions with experimental data in accelerator 
experiments. The main goal of this paper is to compare the fluxes obtained 
with the NLO and with the LO, i.e. we will compute the $K$ factor for the 
neutrino and muon fluxes. This $K$ factor is necessarily different from the $ 
K$ factor for charm production (which can be found in the literature), 
because only the forward going leptons contribute significantly to the 
atmospheric fluxes. 

A similar comparison was very recently made in \cite{PRS}, using the
approximate analytical solutions introduced by TIG to the cascade equations in
the atmosphere.  We make instead a full simulation of the cascades, using the
combined MNR and PYTHIA programs. These two treatments of the problem are
complementary. For comparison, we include results obtained with the CTEQ 3M
gluon structure function used in Ref.~\cite{PRS}. We find our CTEQ 3M results
to be close to those of the PRS study, in spite of the very different
approaches used in the two calculations.
 
Addressing right away a concern that has been expressed to us several times,
about the applicability of perturbative QCD calculations, mostly done for
accelerator physics, to the different kinematic domain of cosmic rays, we would
like to point out that, since the characteristic charm momentum in our
simulations is of the order of the charm mass, $k\simeq O(m_{c})$, we do not
have here
the uncertainty present in the differential cross sections \cite{MNR}, when $ 
k_{T}$ is much larger than $m_{c}$ (as is the case in accelerators), due to 
the presence of large logarithms of $(k_{T}^{2}+m_{c}^{2})/m_{c}^{2}$. 
Depending on the steepness of the gluon structure function we take, we do 
have, however, large logarithms, known as ``ln(1/x)'' terms, where $x\simeq 
\sqrt{4m_{c}^{2}/s}$ ($s$ is the hadronic center of mass energy squared) is 
the average value of the hadron energy fraction needed to produce the $c\bar{ 
c}$ pair. These should not be important for steep enough gluon structure 
functions (namely for values of $\lambda$ in Eq. (9) not very close to 
zero), but we have not made any attempt to deal with this issue. 
 
In the next section of this paper we explain our normalization of the NLO 
charm production cross section in the MNR program. In Sect.~\ref 
{sect:simulation} we describe the computer simulations used to calculate the 
neutrino and muon fluxes. In Sect.~\ref{sect:fluxes} we show the results of 
our simulations, we discuss the differences between a NLO and a LO approach 
and we make a comparison with the fluxes of the TIG model. 
 
In this paper we consider only vertical showers for simplicity (the same was 
done by TIG). 
 
\section{Charm production in perturbative QCD} 
 
\label{sect:NLO} 
 
In this section, we show evidence that perturbative QCD gives a fair 
description of the present accelerator data on open charm production in the 
kinematic region most important for cosmic ray collisions in the atmosphere. 
 
There are still not many experiments on open charm production with good 
enough statistics, despite the recent improvements, but many are expected in 
the near future. 
 
We use a NLO approach which is based on the MNR calculation, for which we 
have obtained the computer code. The NLO cross section for charm production 
depends on the choice of the parton distribution functions (PDFs) and on 
three parameters: the charm quark mass $m_{c}$, the renormalization scale $ 
\mu_{R}$, and the factorization scale $\mu_{F}$. 
 
\subsection{Choice of $m_{c}, \protect\mu_{R}, \protect\mu_{F}$} 
 
MNR have two default choices of $m_{c}$, $\mu_{R}$ and $\mu_{F}$: for total 
cross sections they choose $m_{c}$ = 1.5 GeV, $\mu_{R}= m_{c}$, $\mu_{F} = 2 
m_{c}$; for differential cross sections they choose instead $m_{c}$ = 1.5 
GeV, $\mu_{R}= m_{T}$, $\mu_{F} = 2 m_{T}$, where $m_{T} = \sqrt{k_{T}^{2} + 
m_{c}^{2}}$ is the transverse mass. 
 
The current procedure to reproduce the measured differential cross sections  
\cite{alves2, ada, fmnr} is to use the MNR default choices for these three 
parameters and multiply the result by the global factor of about 2 or 3 
necessary to match the predicted and measured total inclusive cross 
sections. Although this procedure might be acceptable in face of the 
uncertainties in the pQCD predictions, we find it unsatisfactory from a 
theoretical point of view. We prefer to fit the differential and total cross 
sections with one and the same combination of $m_{c}$, $\mu_{R}$, and $ 
\mu_{F}$. 
 
We make separate fits of $m_{c}$, $\mu_{R}$, and $\mu_{F}$ for each of the 
following sets of PDFs: MRS R1, MRS R2 \cite{MRS1}, CTEQ 3M \cite{CTEQ3} and 
CTEQ 4M \cite{CTEQ} (see the next subsection for details). 
 
We are aware that several choices of $m_{c}$, $\mu_{R}$ and $\mu_{F}$ may 
work equally well. In fact the cross sections increase by decreasing $\mu_{F} 
$, $\mu_{R}$ or $m_{c}$, so changes in the three variables can be played 
against each other to obtain practically the same results. We present here 
just one such choice. 
 
We choose $\mu_{R} = m_T$, $\mu_{F} = 2 m_T$ for all sets, and  
\begin{eqnarray} 
m_{c}&=&1.185{\rm ~GeV} \quad\hbox{for MRS R1,}  \label{eq:1} \\ 
m_{c}&=&1.31{\rm ~GeV} \quad\hbox{for MRS R2,}  \label{eq:1b} \\ 
m_{c}&=&1.24{\rm ~GeV} \quad\hbox{for CTEQ 3M,}  \label{eq:1c} \\ 
m_{c}&=&1.27{\rm ~GeV} \quad\hbox{for CTEQ 4M.}  \label{eq:1d} 
\end{eqnarray} 
 
We fit $m_{c}$, $\mu_{R}$, and $\mu_{F}$ to the latest available data on 
charm production \cite{alves1, alves2, ada, fmnr} in proton-nucleon and 
pion-nucleon collisions. We use mainly the data on $pN$ collisions, which 
are more relevant to us, but examine also the $\pi N$ data to see how well 
our choice of parameters works there. 
 
The MNR program calculates the total cross section for $c\bar c$ pair 
production, $\sigma_{c \bar{c}}$. We converted the experimental data on $ 
D^{+}$ or $D^{-}$ production $\sigma(D^{+},D^{-})$, $D^{0}$ or ${\bar{D}}^{0} 
$ production $\sigma (D^{0},{\bar{D}}^{0})$, or the same cross sections just 
for $x_{F} > 0$ ($x_{F}$ is the Feynman $x$), $\sigma_{+} (D^{+},D^{-})$ and  
$\sigma_{+}(D^{0},\bar D^{0})$, into $\sigma_{c\bar c}$ values following  
\cite{fmnr}. 
 
The data we used for the `calibration' of the MNR program are shown in
Table~\ref{table:1} and Table~\ref{table:2} \cite{alves1, alves2, ada, fmnr}.
These tables also present a comparison of experimental data on total inclusive
D-production cross sections (converted to $\sigma_{c\bar c}$ total cross
sections) with those calculated with the MNR program.
 
For the data of Table~\ref{table:1}, for $pN$ collisions, the conversion is 
done using    
\begin{equation} 
\sigma_{c\bar{c}}=1.5\times{\frac{1}{2}}\times\lbrack{\sigma(D^{+},D^{-})+ 
\sigma(D^{0},\bar{D}^{0})]}   \label{eq:2} 
\end{equation} 
if cross sections are measured for any $x_{F}$, or    
\begin{equation} 
\sigma_{c\bar{c}}=1.5\times2\times{\frac{1}{2}}~~[{\sigma_{+}(D^{+},D^{-})+ 
\sigma_{+}(D^{0},\bar{D}^{0})]}~,   \label{eq:3} 
\end{equation} 
if experimental data are given for $x_{F}>0$ only. The explanation of the 
factors in Eqs. (\ref{eq:2}),(\ref{eq:3}) is as follows. The $\frac{1}{2}$ 
factors convert single $D$ inclusive into ${D\bar{D}}$ pair inclusive cross 
sections. The 1.5 factors are required to take into account the production 
of $D_{S}$ and $\Lambda_{c}$ (which is included in $\sigma_{c\bar{c}})$ 
through the ratios \cite{fmnr}    
\begin{equation} 
{\frac{\sigma(D_{S})}{{\sigma(D^{+},D^{0})}}}\simeq0.2,\qquad{\frac 
{\sigma(\Lambda_{c})}{{\sigma(D^{+},D^{0})}}}\simeq0.3,   \label{eq:4} 
\end{equation} 
(the same relation also for antiparticles). The factor 2 in Eq. (\ref{eq:3}) 
converts from $x_{F}>0$ to all $x_{F}$ (i.e. it is $\sigma_{c\bar{c} 
}/\sigma_{c\bar{c}}(x_{F}>0)$ for the $pN$ case). 
 
In the case of $\pi N$ collisions (Table~\ref{table:2}) the factor 2 in 
equation (\ref{eq:3}) is replaced by 1.6, which is the value of $ 
\sigma_{c\bar c}/\sigma_{c\bar c}(x_{F} > 0)$ when a pion beam is used. 
 
Table~\ref{table:1} explains our choice of $m_{c}$ values. The $m_{c}$ 
values in Eqs.(\ref{eq:1}),(\ref{eq:1b}),(\ref{eq:1c}) and (\ref{eq:1d}) 
reproduce well the central values of the $pN$ charm inclusive total cross 
sections \cite{alves1}, using the program with the four different PDFs. 
 
In Table~\ref{table:2} we also present a similar analysis for $\pi N$ 
collisions, using only MRS R1 for simplicity. In this case slightly higher 
values of $m_{c}$ fit the $\pi N$ data \cite{alves1, fmnr} a bit better, 
while $m_{c} = 1.185$ GeV, the value we take with the MRS R1 PDF, fits the $ 
pN$ data \cite{alves1, alves2, fmnr} a bit better. Notice that for the pions 
we used a different PDF, SMR2 \cite{harri}, the same used in Refs. \cite 
{alves1, alves2} (obviously not used in our calculations of atmospheric 
fluxes). We present the $\pi N$ data just for completeness, to show that 
they too are reasonably well fitted with our choice of parameters. These 
other values of $m_{c}$ in Table~\ref{table:2} well reproduce the $\pi^{\pm}N 
$ data at 250 GeV \cite{alves1} and the $\pi^{-}N$ data at 350 GeV \cite{ada} 
(which seem a bit too low with respect to the data at 250 GeV). Even if each 
value of $m_{c}$ reproduces best each total cross section, all three provide 
reasonable fits to all data, as can be seen also in the Figs.~1--3. 
 
In Figs.~1--3 we present total and differential cross sections calculated 
with the MNR program and compared to the experimental data. As a way of 
example,  we describe our fits for MRS R1 only. 
 
Fig.~1a shows the fit to $pN$ total cross sections (converted into $ 
\sigma_{c\bar c}$ values as described above). In addition to the 
experimental value of Table~\ref{table:1} --- which is the fundamental one, 
since it's the experiment whose differential cross sections we want also to 
fit --- we added other experimental points coming from previous experiments 
(for details see \cite{fmnr}). For $pN$ the $m_{c}= 1.185$ GeV is the best 
choice. 
 
Fig.~1b shows the same for $\pi N$ collisions. Here, as explained before, 
values of $m_{c}= 1.25$ GeV or $m_{c}= 1.31$ GeV are a better choice. Again 
we added here for completeness other experimental points coming from 
previous experiments \cite{fmnr}. 
 
Fig.~2ab shows fits to D-inclusive differential cross sections. In this 
figure the theoretically obtained $d\sigma_{c\bar c}/dx_{F}$ and $ 
d\sigma_{c\bar c}/dp_{T}^{2}$ were converted into D-cross sections, with no 
extra factors. Fig.~2ab presents the data of the E769 collaboration \cite 
{alves2} for $pN$ and $\pi N$ at 250 GeV. In these cases the differential $ 
\sigma_{c\bar c}$ cross sections are converted into single inclusive ones 
(by a factor of 2) and then into cross sections for production of $D^{\pm}, 
D^{0}, \bar D^{0}$ and $D_{S}^{\pm}$ (by a factor of 1.2/1.5, see Eq.~(\ref 
{eq:4})) for the E769 data. For example, 
 
\begin{equation} 
{\frac{d\sigma}{dx_{F}}}(D^{\pm}, D^{0},\bar{D^{0}}, D_{S}^{\pm}) ~\simeq{ 
\frac{1.2}{1.5}} \times2 \times{\frac{d\sigma_{c\bar c}}{dx_{F}}}  
\label{eq:5} 
\end{equation} 
for Fig.~2a (and similar factors for $d\sigma/dp_{T}^{2}$ for Fig.~2b). The 
fit to the $d\sigma/dp_{T}^{2}~~ pN$ data in Fig.~2b seems to be a bit too 
low, but it is not very different from the fit shown in Fig.~2 of reference  
\cite{alves2}. The predicted $d\sigma/dp_{T}^{2}$ are not sensitive to 
differences in $m_{c}$ that are instead more noticeable in $d\sigma/dx_{F}$. 
 
Fig.~3ab presents the $\pi N$ data at 350 GeV of the WA92 collaboration \cite 
{ada} in a way similar to Fig.~2ab. In these cases the differential $ 
\sigma_{c\bar c}$ cross sections are converted into a single inclusive ones 
(by a factor of 2) and then into cross sections for production of $D^{\pm}$,  
$D^{0}$ and $\bar D^{0}$ only (by a factor of 1.0/1.5, see Eq.~(\ref{eq:4})) 
for the WA92 data. Similar conclusions can be drawn: for pions $m_{c}= 1.31$ 
GeV is the best choice in this case. 
 
We have performed the same analysis with MRS R2, CTEQ 4M and CTEQ 3M, even 
if we do not show here any of the fits. The results for total and 
differential cross sections were similar to those shown for the MRS R1, the 
only difference being the choice of $m_{c}$. 
 
In conclusion, we obtain good fits to all data on charm production with one 
choice of $\mu_{R},\mu_{F}$ and $m_{c}$ for each PDF, without other 
normalizations. 
 
\subsection{Choice of PDFs} 
 
Consider the collision of a cosmic ray nucleus of energy $E$ per nucleon, 
with a nucleus of the atmosphere in which charm quarks of energy $E_{c}$ are 
produced, which decay into leptons of energy $E_{l}$ (in the lab. frame, 
namely the atmosphere rest frame). Due to the steep decrease with increasing 
energy of the incoming flux of cosmic rays, only the most energetic charm 
quarks produced count for the final lepton flux, and these $c$ quarks come 
from the interactions of projectile partons carrying a large fraction of the 
incoming nucleon momentum. Thus, the characteristic $x$ of the projectile 
parton, that we call $x_{1}$, is large. It is $x_{1}\simeq O(10^{-1})$. We 
can, then, immediately understand that very small parton momentum fractions 
are needed in our calculation, because typical partonic center of mass 
energies $\sqrt{\hat{s}}$ are close to the $c\bar{c}$ threshold, $ 
2m_{c}\simeq 2$ GeV (since the differential cross section decreases with 
increasing ${\hat{s}}$), while the total center of mass energy squared is $ 
s=2m_{N}E$ (with $m_{N}$ the nucleon mass, $m_{N}\simeq 1$ GeV). Calling $ 
x_{2}$ the momentum fraction of the target parton (in the nuclei of the 
atmosphere), then, $x_{1}x_{2}\equiv \hat{s}/s=4m_{c}^{2}/(2m_{N}E)\simeq $ 
GeV/$E$. Thus, $x_{2}\simeq O$(GeV/0.1 $E$), where $E$ is the energy per 
nucleon of the incoming cosmic ray in the lab. frame. The characteristic 
energy $E_{c}$ of the charm quark and the dominant leptonic energy $E_{l}$ 
in the fluxes are $E_{l}\simeq E_{c}\simeq 0.1E$, thus $x_{2}\simeq O$(GeV/ $ 
E_{l}$), as mentioned above. 
 
For $x>10^{-5}$ ($E\lesssim10^3$ TeV), PDFs are available from global 
analyses of existing data. We use four sets of PDFs. MRS R1, MRS R2 \cite 
{MRS1} and CTEQ 4M \cite{CTEQ}, incorporate most of the latest HERA data and 
cover the range of parton momentum fractions $x\geq10^{-5}$ and momentum 
transfers $Q^{2}\geq1.25-2.56$ GeV$^{2}$. MRS R1 and MRS R2 differ only in 
the value of the strong coupling constant $\alpha_{s}$ at the Z boson mass: 
in MRS R1 $\alpha_{s}(M_{Z}^{2})=0.113$, and in MRS R2 $ 
\alpha_{s}(M_{Z}^{2})=0.120$. The former value is suggested by ``deep 
inelastic scattering'' experiments, and the latter by LEP measurements. This 
difference leads to different values of the PDF parameters at the reference 
momentum $Q_{0}^{2}=1.25\ {\rm GeV}^{2}$ where the QCD evolution of the MRS 
R1 and R2 PDFs is started. The CTEQ 4M is the standard choice in the $ 
\overline{MS}$ scheme in the most recent group of PDFs from the CTEQ group ($ 
\alpha_{s}(M_{Z}^{2})=0.116$ for CTEQ 4M). We also use an older PDF by the 
CTEQ group, namely the CTEQ 3M \cite{CTEQ3}, only for comparisons with \cite 
{PRS}, where it is used as the main PDF. 
 
For $x<10^{-5}$ ($E\gsim 10^3$ TeV), we need to extrapolate the available 
PDFs. For $x\ll1$, all these PDFs go as  
\begin{equation} 
xf_{i}(x,Q^{2})\simeq A_{i}x^{-\lambda_{i}(Q^{2})},   \label{eq:7} 
\end{equation} 
where $i$ denotes valence quarks $u_{v},d_{v}$, sea quarks $S$, or gluons $g$ 
. The PDFs we used (except the older CTEQ 3M) have $\lambda_{S}(Q_{0}^{2}) 
\not =\lambda_{g}(Q_{0}^{2})$, in contrast to older sets of PDFs which 
assumed an equality. As $x$ decreases the density of gluons grows rapidly. 
At $x\simeq0.3$ it is comparable to the quark densities but, as $x$ 
decreases it increasingly dominates over the quark densities, which become 
negligible at $x\lsim 10^{-3}$. 
 
We need, therefore, to extrapolate the gluon PDFs to $x<10^{-5}$. 
Extrapolations based on Regge analysis usually propose $xg(x)\sim 
x^{-\lambda }$ with $\lambda \simeq 0.08$ \cite{lowx}, while evolution 
equations used to resum the large logarithms $\alpha _{s}\ln (1/x)$ 
mentioned above, such as the BFKL (Balitsky, Fadin, Kuraev, Lipatov \cite 
{BFKL}) find also $xg(x)\sim x^{-\lambda }$ but with $\lambda \simeq 0.5\ $ 
\cite{lowx}. A detailed analysis of the dependence of the neutrino fluxes on 
the low $x$ behavior of the PDFs will be given in another publication \cite 
{GGVlambda}. As mentioned above, in the present paper our goal is to compare 
NLO to BORN simulations, for which we use a simplified extrapolation at low $ 
x$ of the gluon PDF, which is somewhat in between the two extreme 
theoretical behaviors described above. For MRS R1-R2 and CTEQ 4M we take a 
linear extrapolation of $\ln g(x)$ as a function of $\ln x$, in which we 
took $\ln g(x)=-(\lambda _{g}(Q^{2})+1)\ln x+\ln A_{g}$, where $\lambda 
_{g}(Q^{2})$ was taken as its value at $x=10^{-5}$, the smallest $x$ for 
which the PDFs are provided; for the CTEQ 3M we used a polynomial 
approximation which is included in the PDF package. 
 
\section{Simulation of particle cascades in the atmosphere} 
 
\label{sect:simulation} 
 
We simulate the charm production process in the atmosphere and the subsequent
particle 
cascades, by modifying and combining together two different programs: the 
MNR routines \cite{MNR} and PYTHIA 6.115 \cite{jetset}.
 
The MNR program was modified to become an event generator for charm 
production at different heights in the atmosphere and for different energies 
of the incoming primary cosmic rays. 
 
The charm quarks (and antiquarks) generated by this first stage of the 
program are then fed into a second part which handles quark showering, 
fragmentation and the interactions and decays of the particles down to the 
final leptons. The cascade evolution is therefore followed throughout the 
atmosphere: the muon and neutrino fluxes at sea level are the final output 
of the process. 
 
In this section we give a brief description of the main parts of the 
simulation. Even if our program is completely different from the one used by 
TIG, because it is constructed around the MNR main routines, nevertheless we 
keep the same modeling of the atmosphere and of the primary cosmic ray flux 
as in TIG and the same treatment of particle interactions and decays in the 
cascade. 
 
Our main improvement is the inclusion of a true NLO contribution for charm 
production (and updated PDFs), so we keep all other assumptions of the TIG 
model in order to make our results comparable to those of TIG. We study the
effect of modifying some of their other assumptions elsewhere \cite{GGVlambda}.
 
\subsection{The model for the atmosphere} 
 
We assume a simple isothermal model for the atmosphere. Its density at 
vertical height $h$ is  
\begin{equation} 
\rho(h)={\frac{X_{0}}{h_{0}}}\,e^{-h/h_{0}},   \label{eq:8} 
\end{equation} 
where the scale height $h_{0}=6.4\ {\rm km}$ and the column density $
X_{0}=1300\ {\rm g/cm^{2}}$ at $h=0$ are chosen as in TIG, to fit the actual
density in the range $3\ {\rm km}<h<40\ {\rm km,}$ important for cosmic ray
interactions. Along the vertical direction, the amount of atmosphere traversed
by a particle, the depth $X$, is related to the height $h$
simply by   
 \begin{equation} 
X=\int_{h}^{\infty}\rho(h^{\prime})dh^{\prime}=X_{0}e^{-h/h_{0}}.  
\label{eq:9} 
\end{equation} 
The atmospheric composition at the important heights is approximately 
constant: 78.4\% nitrogen, 21.1\% oxygen and 0.5\% argon with average atomic 
number $\langle A\rangle$ = 14.5. 
 
\subsection{The primary cosmic ray flux} 
 
Following TIG \cite{TIG}, we neglect the detailed cosmic ray composition and 
consider all primaries to be nucleons with energy spectrum  
\begin{equation}  
  \phi_N (E, 0) \left [\rm{nucleons}\over {cm^2~ s ~sr ~GeV~/A} \right ]  
  =  
\begin{cases}  
  {1.7 (E/{\rm GeV})^{-2.7} & for $ E < 5~10^6$ GeV\cr 
   174 (E/{\rm GeV})^{-3} & for $ E > 5~10^6$ GeV \cr}  
\end{cases}  
\label{eq:10}  
\end{equation}
The primary flux is attenuated as it penetrates into the atmosphere by 
collisions against the air nuclei. An approximate expression for the 
intensity of the primary flux at a depth $X$ is (see \cite{TIG} again)  
\begin{equation} 
\phi_{N}(E,X)=e^{-X/\Lambda_{N}(E)}~\phi_{N}(E,0)~.   \label{eq:11} 
\end{equation} 
The nuclear attenuation length $\Lambda_{N}$, defined as  
\begin{equation} 
\Lambda_{N}(E)={\frac{\lambda_{N}(E)}{1-Z_{NN}(E)}}~,   \label{eq:12} 
\end{equation} 
has a mild energy dependence through $Z_{NN}$ and $\lambda_{N}$. Here $Z_{NN}$
is the spectrum-weighted moment for nucleon regeneration in nucleon-nucleon
collisions, for which we use the values in Fig.~4 of Ref. \cite{TIG}. And
$\lambda_{N}$ is the interaction thickness
\begin{equation} 
\lambda_{N}(E,h)={\frac{\rho(h)}{\sum_{A}\sigma_{NA}(E)n_{A}(h)}}~,  
\label{eq:13} 
\end{equation} 
where $n_{A}(h)$ is the number density of air nuclei of atomic weight $A$ at
height $h$ and $\sigma_{NA}(E)$ is the total inelastic cross section for
collisions of a nucleon $N$ with a nucleus $A$.\footnote{We recall that the
  elastic cross section contributes negligibly to the primary flux attenuation
  because the average elastic energy loss is very small, less than 1 GeV at the
  high energies we consider. This can be seen using the differential elastic
  cross section $d\sigma_{el}/dQ^{2}=(d \sigma_{el}/dQ^{2})_{Q^{2}=0}\exp
  (-bQ^{2})$ with $b=[7.9+0.9\ln p_{lab}]{\rm GeV}^{-2}$, with $p_{lab}$ in GeV
  \cite{goulianos}. Here $Q$ is the momentum transfer of the colliding proton
  of incoming momentum $p_{lab}$ and mass $M$. The mean energy loss is the mean
  value of $Q^{2}/2M$ (here $M$ is the target proton mass) namely
  $(1/2Mb)=67{\rm MeV}/(1+0.1\ln(p_{lab}/{\rm GeV}))$. This is 46 MeV at
  $E=100{\rm GeV}$, and smaller at higher energies.} This cross section scales
essentially as $A^{2/3}$, since for the large nucleon-nucleon cross sections we
deal with, the projectiles do not penetrate the nucleus. So we set
$\sigma_{NA}(E)=A^{2/3}\sigma_{NN}(E)$. For $\sigma_{NN}(E)$ we use the fit to
the available data in Ref. \cite{RPP96}.  Using our height independent
atmospheric composition, we simplify Eq.~(\ref{eq:13}) as follows,
\begin{equation} 
\lambda_{N}(E,h)={\frac{\langle A\rangle}{\langle A^{2/3}\rangle}}\,{\frac{ 
{\rm u}}{\sigma_{NN}(E)}}=2.44\,{\frac{{\rm u}}{\sigma_{NN}(E)}}~.  
\label{eq:14} 
\end{equation} 
Here $\langle~\rangle$ denotes average and u is the atomic mass unit, that 
we write as  
\begin{equation} 
{\rm u}=1660.54{\rm ~mb~g/cm^{2}}.   \label{eq:14b} 
\end{equation} 
We therefore find that in our approximations $\lambda_{N}(E)$ is independent 
of height. 
 
\subsection{Charm production with MNR routines} 
 
As we remarked before, the modified MNR routines are the first stage of our
simulation. For a given energy $E$ of a primary incoming proton in the lab
system, i.e.\ in the atmosphere reference frame, we generate a collision with a
nuclear target at rest in the atmosphere, activating the MNR routines (primary
event, $pN$ collision, with $N=(p+n)/2$) .
 
These routines generate total and differential cross sections through a 
VEGAS integration, which creates a large number of `subevents', each one 
with a particular weight, which in the original MNR program are summed 
together to calculate the final cross sections. 
 
It is easy to modify the program so that each of these subevents (together with
its weight) can represent the production of a charm $c$ (or of a $c \bar{c}$
pair, or $c \bar{c}$ gluon, etc.) with given kinematics in any particular
reference frame of interest. The original MNR routines can calculate single
differential cross sections, in which the kinematics of only one final $c$
quark is available, and double differential cross sections, in which the full
kinematics of the $c \bar{c}$ pair (plus an additional parton in NLO processes)
becomes available, for each subprocess. We have used both these possibilities.
We will refer to them as `single' and `double' modes.  The `single' is the mode
we use to obtain all our results. We use the `double' mode only to compare the
results of the independent fragmentation model used in the evolution of
cascades in the `single' mode, with the more reliable string fragmentation
model, which can only be used in the `double' mode, as we explain below.
 
The MNR program \cite{MNR, nde} contains all BORN and NLO processes. In the
`single' mode we can generate the following processes, with only the kinematics
of the $c$ quark available,
\begin{equation} 
gg\rightarrow cX;\;\;q\bar{q}\rightarrow cX\;({\rm BORN})\quad gg\rightarrow 
cX;\;\;q\bar{q}\rightarrow cX;\;\;qg\rightarrow cX\;({\rm NLO})   \label{eq:15} 
\end{equation} 
where q represents any light quark or antiquark. In the `double' mode we have the following 
processes  
\begin{equation} 
gg\rightarrow c\bar{c};\;\;q\bar{q}\rightarrow c\bar{c}\;({\rm BORN})\quad 
gg\rightarrow c\bar{c}g;\;\;q\bar{q}\rightarrow c\bar{c}g;\;\;qg\rightarrow c 
\bar{c}q\;({\rm NLO})   \label{eq:16} 
\end{equation} 
for which the kinematics of all the outgoing partons is fully determined for 
each `subevent'. 
 
All the kinematical variables of the partons in the final state constitute 
the input for the next stage of the program, described in the next 
subsection. 
 
An important characteristic of the first stage is that, besides $m_c$, $\mu_R$,
and $\mu_F$, we can select any desired PDF to be used with the charm production
routines. We have updated the set of PDFs in the original MNR program.
 
According to the discussion of Sect.~\ref{sect:NLO}, we use the MRS R1, MRS R2,
CTEQ 3M and CTEQ 4M parton distribution functions, together with the values of
$m_c$, $\mu_R$, and $\mu_F$ in Eqs.~(\ref{eq:1}--\ref{eq:1d}).
 
As a concrete example of the integrals performed in our program, here we 
write the differential flux $\phi_{\mu}$ of muons (namely of $\mu^{+}+\mu^{-} 
$) with energy $E_{\mu}$ \ ($\mu$ stands here for $\mu^{+}$ or $\mu ^{-}$) 
in the `single' mode ($\phi_{\mu}$ has units cm$^{-2}$ s$^{-1}$ sr$^{-1}$ GeV 
$^{-1}$)  
\begin{eqnarray}  
  \phi_\mu(E_\mu) & = &\int_{E_\mu}^{\infty} dE \int_0^{\infty} dh\  
  \phi_N(E,X(h)) \sum_A n_A(h) \times \nonumber \\ &&\int_{E_\mu}^{E} dE_c 
  \left [{d \sigma(p A \to c Y;E,E_c) \over d E_c} \right]_{MNR} \left [ 
    {dN_\mu (c \to \mu; E_c, E_\mu, h) \over d E_{\mu} } \right]_{PYTHIA} 
  \nonumber \\  &+& (c \to \bar{c}). 
 \label{eq:flux1}  
\end{eqnarray}  
Here $n_{A}(h)$ is the number density of nuclei of atomic number A in the 
atmosphere, $E$ is the energy of the primary cosmic ray proton, $E_{c}$ the 
energy of the charm produced in the collision $pA\rightarrow cY$ \ ($Y$ here 
stands for anything else). Using the relation ${d\sigma(pA\rightarrow 
cY)/dE_{c}}=A\ {d\sigma(pN\rightarrow cY)/dE_{c}}$, the sum over $A$ becomes  
$\sum_{A}n_{A}(h)A=\rho(h)/u$. Using $dX=-\rho(h)dh$, Eq. (\ref{eq:11}), and 
normalizing to one the distribution in depth $X$, $\phi_{\mu}$ becomes  
\begin{equation} 
\phi_{\mu}(E_{\mu})=\int_{E_{\mu}}^{\infty}dE\int_{X_{0}}^{\infty}dX\ \phi 
_{N}(E,X\mathord{=}0)\ {\frac{e^{{-X/\Lambda}_{N}{(E)}}}{\Lambda_{N}(E)}\ } 
\left[ {\frac{f(h)\Lambda_{N}(E)}{u}}\right] ,   \label{eq:flux2} 
\end{equation} 
where, from Eqs.(\ref{eq:12}) and (\ref{eq:14}), $\Lambda_{N}/u=2.44[\sigma 
_{NN}(1-Z_{NN})]^{-1}$ and  
\begin{equation} 
f(h)=2\int_{E_{\mu}}^{E}dE_{c}\left[{\frac{d\sigma(pN\rightarrow cY;E,E_{c}) 
}{dE_{c}}}\right]_{MNR}\ \left[\frac{dN_{\mu}(c\rightarrow 
\mu;E_{c},E_{\mu},h)}{dE_{\mu}}\right]_{PYTHIA}
 \label{eq:flux3} 
\end{equation} 
Here the factor of 2 accounts for the muons produced by ${\bar{c}}$ (only c 
quarks are used in the program for simplicity); the $pN$ inclusive charm 
production cross section is computed with the MNR program (here are the 
integrations over the PDFs and partonic cross sections) and the last square 
bracket is the number of muons of energy $E_{\mu}$ which reach sea level, 
produced in the cascades simulated by PYTHIA. Each cascade is initiated by a  
$c$ quark (in the `single' case) of energy $E_{c}$ and momentum $k$ 
(provided by the MNR routines) at a height $h$ chosen through a random 
number $R$ homogeneously distributed between 0 and 1, which gives the value 
of the $X$ probability distribution in Eq. (\ref{eq:flux2}), namely $R=e^{{ 
-X/\Lambda }_{N}{(E)}}$. 
 
\subsection{Cascade evolution with PYTHIA routines} 
 
The parton $c$ (or partons in the `double' case) generated by the first 
stage, namely by the MNR routines, are entered in the event list of PYTHIA 
and they become the starting point of the cascade generation. 
 
PYTHIA first fragments the $c$ quark (in the `single' mode, or all the 
partons in the `double' mode) into hadrons, after showering, which can be 
optionally shut off. The charm quarks hadronize into $D^{0}$, $\bar{D}^{0}$,  
$D^{\pm}$, $D_{s}^{\pm}$ and $\Lambda_{c}$. We used here the Peterson 
fragmentation function option. For each hadron produced, a simple routine 
added to PYTHIA decides if the hadron interacts in the atmosphere (loosing 
some energy) or decays. This is the same approach as in TIG. PYTHIA follows 
in this way the cascade in the atmosphere and populates the histograms of 
muons and neutrinos as a function of their different energies. We mention 
here a few important technical details. The `single' and `double' modes 
described before use different fragmentation models. In the `single' mode 
only one $c$ quark is available and is entered at the beginning of the event 
list (with its energy and momentum in the partonic CM reference frame). In 
this case PYTHIA uses the `independent fragmentation' model (see \cite 
{jetset} for details). We only include $c$ quarks and at the end multiply 
the result by a factor of two to account for initial $\bar{c}$ quarks. 
 
In the `double' mode, instead, which we only use at the LO, we start with 
two ($c\bar{c}$) partons in the event list. In this case we opt to use the 
`string fragmentation' model (Lund model, \cite{jetset}). This model 
generally gives better results than the independent fragmentation, in which 
energy and momentum conservation have to be imposed a posteriori and whose 
results depend on the reference frame used, which empirically is chosen to 
be the partonic CM frame. To impose energy and momentum conservation in the 
independent fragmentation,  we used the option (MSTJ(3)=1, see again \cite 
{jetset}) in which particles share momentum imbalance compensation according 
to their energy (roughly equivalent to boosting events to CM frame) but we 
have convinced ourselves that the results do not depend much on the way of 
imposing energy and/or momentum conservation, because trial runs with 
different options have given similar results for the fluxes. 
 
Even if independent fragmentation is in general less desirable than string 
fragmentation, we use the `single' mode as our main choice.  The main reason 
to use the `single' mode is that the simulations run in acceptably short 
times (a few days)  on the SUN computers we use, while giving results 
practically identical to the `double' mode in the comparisons we have made 
(see Fig.~6c). The simulation of the cascades in the `double' mode takes 
between  five and ten times longer. We tested the goodness of the 
independent fragmentation by comparing the outcome of fluxes computed at the 
Born level, in which the charm fluxes at production are identical (we put 
one $c$ in the atmosphere and multiply the outcome by two to account for the  
${\bar c}$ in one case, and we put $c {\bar c}$ in the atmosphere, instead, 
in the second case)  and the sole difference in both modes is due to the 
different fragmentation models used. The results were extremely close (at 
Born level the difference is less than $5\%$, at energies above $10^5 GeV$), 
as can be seen in Fig.~6c. 
 
Apart from the mentioned differences between the `single' and `double' 
modes, the simulations then proceed basically in the same way in both modes. 
For each of the `subevents', i.e. for each set of initial parton(s) put in 
the event list, a certain height in the atmosphere is randomly chosen as 
explained above, this being the position at which the partons are generated 
from the initial proton-nucleon collision. This random height $h$ is 
generated in a way similar to TIG (see Ref. \cite{TIG}), but different, 
because we include a correction for nucleon regeneration in nucleon-nucleon 
collisions by using $\Lambda_{N}$, the nuclear attenuation length, in Eq. (\ref{eq:11}) instead of $\lambda_{N}$ , the interaction thickness (see Eqs.~(\ref{eq:12}),(\ref{eq:13}) and (\ref{eq:14})).The only difference compared 
to TIG (see Eq. (15) in the last paper of Ref. \cite{TIG}) is the inclusion 
of the $(1-Z_{NN})$ correction term. This was done because we could not 
include regenerated protons directly in our simulation of the cascades, 
since events and subevents are now created by the MNR routines and not by 
PYTHIA, as it was in TIG. 
 
When parton showering is included at the beginning of the cascade simulation 
performed by PYTHIA,  some double counting is present. The double counting 
appears when a LO diagram, for example $gg \to c \bar{c}$, with a subsequent 
splitting contained in PYTHIA, for example $c \to g c$ is summed to NLO 
diagram, $gg \to g c \bar{c}$ with the same topology, as if both diagram 
were independent, when actually the NLO contains the first contribution when 
the intermediate $c$ quark on mass shell. We have not tried to correct this 
double counting but have instead confronted  the results obtained including 
showering (our standard option) with those excluding showering (in which 
case there is no double counting) and found very similar leptonic fluxes 
(see Fig.~6b). 
 
The particles generated after the initial hadronization are then followed 
throughout the atmosphere and PYTHIA evolves the cascade with the same 
treatment of interactions and decays proposed by TIG. The final number of 
muons and neutrinos at sea level is therefore calculated considering all the 
`subevents', each with its respective weight $W_{i}$ from the MNR program, 
which produce the final particles through all the possible decay channels of 
charmed particles decaying into prompt leptons. Since only the decay modes 
of charmed hadrons going into $\mu$ or $\nu_{\mu}$ or $\nu_{e}$ are left 
open in the simulation, and there are essentially just 2 modes for each 
charmed particle (for example: $D^{+}\rightarrow e^{+\ }\nu_{e}\ +anything$ 
, with branching ratio $=0.172$; $D^{+}\rightarrow\mu^{+\ }\nu_{\mu }\ 
+anything$, with branching ratio $=0.172$; all other channels closed), the 
branching ratios for each of these modes is fictitiously taken by PYTHIA to 
be 1/2 and need to be normalized by multiplying by the actual branching 
ratio ($0.172$ for the example above) and dividing by $1/2$. Besides, since 
not all events are accepted by PYTHIA to generate a complete cascade, the 
result is normalized by dividing by the sum of all the weights of accepted 
events and multiplying it by the total c inclusive cross section. 
 
\subsection{Summary} 
 
To summarize, our computation of the final fluxes is organized as follows. 
 
$\bullet$ An external loop over the primary energy $E$ generates an 
integration over $E$ in the range $10^{1} - 10^{11} GeV$. 
 
$\bullet$ For each primary energy $E$, the MNR routines generate `subevents' 
with weight $W_{i}$, for all the LO and NLO processes. 
 
$\bullet$ Each subevent is assigned a random height (so that implicitly an 
integration over $h$ is performed) and all this is passed to PYTHIA as a 
definite set of parton(s) to be put at the beginning of the event list. 
 
$\bullet$ For each of these `subevents', PYTHIA treats showering 
(in our standard option), hadronization and evolution of the  cascade in the 
atmosphere, and generates the final leptons. 
 
$\bullet$ For each decay channel of interest, the produced leptons are 
weighted with $W_{i}$ and then summed into the final fluxes. 
 
\section{Neutrino and muon fluxes} 
 
\label{sect:fluxes} 
 
Figs.~4-6 show the results of our simulations.  Fig.~4 shows the total
inclusive charm-anticharm production cross sections $\sigma_{c\bar c}$, and the
$K$ factor for $c$ production, namely the ratio between the NLO and Born cross
sections, $K_{c} = \sigma_{c\bar c}^{NLO} / \sigma_{c\bar c}^{Born}$, for the
four PDFs we consider and for TIG.  Fig.~5 shows our main results obtained with
our default choice of options: a `single' mode calculation including the
contributions from all processes in Eq.~(\ref{eq:15}) and with parton showering
included in the cascade simulation performed by PYTHIA). Finally Fig.~6 shows
the relative importance of the processes included in the fluxes and a
comparison of the `single' and `double' modes and of the `on' and `off'
showering options.

In Fig.~4a, the total inclusive charm-anticharm production cross sections
$\sigma_{c\bar c}$ are plotted over the energy range needed by our program, $E
\leq 10^{11}$ GeV, for our four different PDFs. They were calculated using the
MNR program, with the `calibration' described in Sect.~\ref{sect:NLO}, up to
the NLO contribution.  For comparison, we also show the cross section used by
TIG and the Born (LO) contribution for one of the PDFs, MRS R1. We see in the
figure that all our cross sections agree at low energies, as expected due to
our `calibration' at 250 GeV, and are very similar for energies up to
$10^{6}-10^{7}$ GeV. At higher energies they diverge, differing by at most 50\%
at the highest energy we use, $10^{11}$ GeV.  In fact, at energies beyond
$10^{7}$ GeV, the CTEQ 3M cross section becomes progressively larger than the
CTEQ 4M and MRS R2 cross sections, which are very close to each other. The MRS
R1 becomes on the contrary progressively lower than the other three.
 
We see in Fig.~4a that for energies above $10^{4}$ GeV our cross sections are
considerably higher than the one used by TIG. This difference can be traced in
part to the use by TIG of an option of PYTHIA by which the gluon PDF is
extrapolated to $x \leq 10^{-4}$ with $\lambda=0.08$, while all the PDFs we use
have a higher value of $\lambda\simeq0.2-0.3$. And in part to TIG scaling the
LO cross sections obtained with PYTHIA by a constant $K$ factor of 2, while at
large energies the $K$ factor is actually larger than 2 by about 10-15\% (see
Fig.~4b).
 
In Fig.~4b we explicitly show the $K$ factor for $c$ production, namely the 
ratio between the NLO and Born cross sections, $K_{c} = \sigma_{c\bar 
c}^{NLO} / \sigma_{c\bar c}^{Born}$, for our PDFs and for TIG.  All the $ 
K_{c}$ values are around the usually cited value of $2$ for most of the 
intermediate energies, but are larger at the lowest energies and also at the 
highest energies (except for CTEQ 3M),  and they all are within about 15\% of 
each other. 
 
Fig.~5 contains three sets of figures, one for each lepton: $\mu$, $\nu_{\mu} $
and $\nu_e$. The left figure of each set shows the $E^{3}$-weighted vertical
prompt fluxes, for all our PDFs up to NLO (labelled `NLO') and, as an example,
the LO (labelled `BORN') for MRS R1, together with the total fluxes up to NLO
of TIG, both from prompt and conventional sources (dotted lines). The right
part of each set shows the corresponding $K_{l}$ value (where
$l=\mu,\nu_{\mu},\nu_e$), i.e. the ratio of the total NLO flux to the Born flux
of the figure on the left. The figures show that our fluxes are higher than
those of TIG for $E > 10^{3}~{\rm GeV}$. Leaving apart differences in the two
simulations that cannot be easily quantified, this discrepancy can largely be
explained by the different cross sections used by TIG and us: the TIG cross
section is lower than ours for $E > 10^{4}$ GeV. Using a value of $\lambda$
similar to TIG ($\lambda \simeq 0$) at small $x$, we obtain fluxes similar to
those of TIG at energies above $10^6$ GeV \cite {GGVlambda}.
 
In particular, our fluxes are all larger than TIG by factors of 3 to 10 at the
highest energies, what puts our fluxes in the bulk-part of previous estimates
(see Refs. \cite{bnsz,c,gghv,zhv}).  There is an evident dependence of the
fluxes on the choice of PDF. It is remarkable that MRS R2 and CTEQ 4M give very
similar results. Those of the MRS R1 become lower and those of the older CTEQ
3M PDF become higher as the energy increases (both differing by about 30-50\%
at the highest energies with respect to the MRS R2-CTEQ 4M fluxes). This is due
to the intrinsic differences of the PDF packages used and the consequent
different extrapolated values of $\lambda$ at small $x$ or high energies.
 
The CTEQ 3M fluxes were included to compare our results with those of Ref.
\cite{PRS}. We find our CTEQ 3M results to be close to those of Ref. \cite
{PRS}, in spite of the very different approaches used in the two calculations.
Our fluxes lie between the two curves for CTEQ 3M shown in Fig.8 of Ref.
\cite{PRS}, corresponding to different choices of renormalization and
factorization scales. Our fluxes are lower (by 30-40\% at $10^{7}{\rm GeV}$),
than the main CTEQ 3M choice of Ref.~\cite{PRS} (solid line
of their Fig.8), which is calculated using values of $\mu_R$, $\mu_F$ and $m_c$
similar to ours.  Our cross section for charm production, for the CTEQ 3M case,
is essentially equal to the one used in Ref. \cite{PRS} (shown in their Fig.
2), so the discrepancies in the final fluxes are to be explained in terms of
the differences in the cascade treatment. It is very difficult to trace the
reasons for these differences.
 
We also see in the figures that, for each PDF, the fluxes for the 
different leptons are very similar: those for $\nu _{\mu }$ neutrinos and $ 
\nu _{e}$ are essentially the same, those for muons are only slightly lower 
(about $10\%$ less at the energies of interest). Also the $K_{l}$'s don't 
differ much for the three leptons, apart from some unphysical fluctuations 
especially evident at the highest energies. Even if they differ  
the PDF, they all show a similar energy dependence, namely they 
increase at low energies and sometimes at high energies also. This behavior 
is also similar to that of the $K_{c}$ factors in Fig.~4b, but with a weaker 
overall energy dependence, as expected, since the leptons of a given energy 
result from $c$ quarks with a range of higher energies. 
 
The $K_l$ factors are all within the range $2.1-2.5$: they are approximately
$2.2$ for MRS R1, $2.4$ for MRS R2 and CTEQ 4M, and $2.3$ for CTEQ 3M. Thus,
our analysis shows that evaluating the lepton fluxes only at the Born level,
and multiplying them by an overall $K_{l}$ factor of about $2.2-2.4$ (i.e. 10
to 20\% larger than the value of $2$ used by TIG\footnote{ We note that in the
  original TIG model there is no distinction between $K_c$ and $K_l$ factors
  since only the Born level is considered. Their $K=2$ factor is just a
  multiplicative constant which can be considered either a $ K_c$ or a
  $K_l$.}), can be good enough to evaluate the NLO fluxes within about 10\%.
Thus we find the approach used by TIG, who multiplied the LO fluxes obtained
with PYTHIA by two, essentially correct, except for their relatively low $K$
factor and the discrepancies existing even at Born level between our fluxes and
those of TIG. In fact, as we mentioned previously, the differences between our
final results and those of TIG depend mostly on the different total inclusive
$c$ cross sections, which can be traced to the extrapolation of the gluon PDF
at small $x$ rather than to the K factor.  Possible causes of the different
results due to the intrinsic differences of the computer simulations cannot be
easily quantified.
 
In Fig.~6 we address three issues. First, we show that the fluxes can be 
obtained within about 30\% with just the gluon-gluon process. This would  
speed up the simulations and, when using the MNR program,  would give 
(contrary to intuition) higher fluxes than those actually derived from all 
processes. Secondly, we show that the fluxes obtained including or excluding 
showering in the simulation made by PYTHIA  (we included showering in our 
standard options) do not differ significantly. The third issue we deal with 
is the difference between the `single' and `double' modes described in Sect. 
3. We show that at LO the results from a `double' mode calculation coincide 
with those of the much shorter `single' mode, that we use in all our 
calculations. Let us deal with these three issues in turn. 
 
In Fig.~6a we show, for a given PDF, the MRS R1, the relative importance of the
different processes contributing to the final fluxes. The solid line is the
total flux obtained as the sum of all the processes of Eq.~(\ref{eq:15}) and
the dotted line shows the result of only gluon-gluon fusion ($gg$), the sum of
Born ($gg$) and pure NLO (excluding Born) $gg$ processes. Also shown are the
separate contributions only at the Born and at the NLO (excluding LO) of both
$gg$ and quark-antiquark ($q {\bar q}$) fusion, what clearly shows that $gg$
dominates. This is to be expected because the gluon PDF is either much larger
than (for $x < 0.1$) or comparable to (for $x \simeq {\cal O}(0.1)$) the quark
PDFs.  The figure plots the absolute value of the quark-gluon ($qg$) terms
because, for the values of the factorization scale that we employ in our
calculations, these terms are negative. This is due to the way the original MNR
calculation is subdivided into processes. In fact, in the MNR program, a part
of the quark-gluon contribution to the cross sections is already contained in
other processes, and must be subtracted in the processes labelled as $qg$.  The
amount subtracted depends on the factorization scale $\mu_{F}$ and may drive
the $qg$ contribution negative. Roughly speaking, if $\mu_{F}$ is small the
$qg$ term is positive,
otherwise (as in our case) the term is negative. The absolute value of the $
qg$ term is in between the $q\bar{q}$ and the $gg$ terms, what makes 
negative the sum of all the processes different from $gg$. Thus, gluon-gluon 
processes alone give a result slightly larger than the total, by about 30\%. 
 
In Fig.~6b we check the effect of shutting off the showering option 
available in PYTHIA. We study only one specific case, the MRS R1. The 
overall effect is minimal: the exclusion of showering slightly increases the 
energy  of the parent charmed hadrons and therefore causes the final fluxes 
of lepton daughters to move towards higher energies;  the effect is barely 
noticeable and just slightly  more important for the Born fluxes (the 
overall difference is about $5\%$). When showering is included some  double 
counting occurs, whose effect must be smaller than the difference between 
the results with showering on and off (since in this case no double counting 
occurs). 
 
Finally in Fig.~6c we confront the `single' and `double' modes of the 
program, for just one PDF, MRS R1, at Born level. At this level, the 
calculation of the charm flux at production is identical (we obtain the 
fluxes from $c$ and multiply by two at the end to account for the ${\bar c}$ 
in one case, and we obtain the fluxes directly from $c {\bar c}$ in the 
other). So, what is actually compared in the two modes at the Born level is 
the fragmentation model: independent fragmentation in the `single' mode and 
string (Lund) fragmentation in the `double' mode. The results from both 
modes at the Born level are almost identical:  as already remarked the 
difference is less than $5\%$ for energies above $10^6~{\rm GeV}$.  
 
\section{Conclusions} 
 
\label{sect:conclusions} 
 
We have used the actual next-to-leading order perturbative QCD calculations 
of charm production cross sections, together with a full simulation of the 
atmospheric cascades, to obtain the vertical prompt fluxes of neutrinos and 
muons. 
 
Our treatment is similar to the one used by TIG, except for the very 
important difference of including the true NLO contribution, while TIG used 
the LO charm production cross section multiplied by a constant $K$ factor of 
2 to bring it in line with the next-to-leading order values. The main goal 
of this paper is to examine the validity of TIG's procedure by computing the 
ratio of the fluxes obtained with the NLO charm production cross section 
versus those obtained with the LO cross section. 
 
These ratios, the $K_l$ factors are between 2.1 and 2.5 for the different gluon
PDFs in the energy range from $10^2$ to $10^9$ GeV (see Fig.~5). Consequently,
our analysis shows that evaluating the lepton fluxes only at the Born level,
and multiplying them by an overall factor of about $ 2.2-2.4$, slightly
dependent on the PDF, can be good enough to evaluate the NLO fluxes within
about 10\%. Therefore, we find the approach used by TIG (i.e.\ multiplying the
LO fluxes by two) essentially correct, except for their relatively low $K$
factor. We find different lepton fluxes than TIG, but this is mostly due to the
discrepancies, even at Born level, between our charm production cross sections
and TIG's.
 
In fact, the prompt neutrino and muon fluxes found by TIG were lower than 
the lowest previous prediction. We find here instead fluxes in the bulk part 
of those previous predictions. This difference can be traced largely to the 
use by TIG of an option of PYTHIA by which the gluon PDF is extrapolated for  
$x \leq 10^{-4}$ with $\lambda=0.08$,  while all the PDFs in this paper have 
a higher value of $\lambda\simeq0.2-0.3$. Using a value of $\lambda$ similar 
to TIG ($\lambda \simeq 0$) we obtain fluxes similar to those of TIG, at 
energies above $10^6$ GeV \cite{GGVlambda}.

\bigskip\bigskip\bigskip\bigskip
 
{\large \noindent \textbf{Acknowledgements}}
 
\bigskip The authors would like to thank the Aspen Center For Physics, where 
this work was initiated, for hospitality, and M. Mangano and P. Nason for 
the MNR program and helpful discussions. This research was supported in part 
by the US Department of Energy under grant DE-FG03-91ER40662 Task C.

\newpage 
 
\section*{FIGURE CAPTIONS} 
 
\begin{description} 
\item[Fig.~1]  Comparison of experimental data for $\sigma_{c\bar c}$ with 
MNR predictions for different $m_{c}$ values: (a) in $pN$ collisions (\cite 
{fmnr}, Table~\ref{table:1}), (b) in $\pi N$ collisions (\cite{fmnr}, Table~ 
\ref{table:2}) (PDF: MRS R1). 
 
\item[Fig.~2]  Comparison of differential cross sections for $(D^{+}, D^{-}, 
D^{0}, \bar D^{0}, D_{S}^{+}$ and $D_{S}^{-})$ production, calculated using 
MNR at different $m_{c}$ values, with E769 data for $pN$ and $\pi N$ \cite 
{alves2}: (a) $d\sigma/dx_{F}$, (b) $d\sigma/dp_{T}^{2}~(x_{F} > 0)$ (PDF: 
MRS R1). 
 
\item[Fig.~3]  Comparison of differential cross sections for $(D^{+}, D^{-}, 
D^{0}, \bar D^{0})$ production, calculated using MNR at different $m_{c}$ 
values, with WA92 data for $\pi N$ \cite{ada}: (a) $d\sigma/dx_{F}$, (b) $ 
d\sigma/dp_{T}^{2}~(x_{F}>0)$ (PDF: MRS R1). 
 
\item[Fig.~4]  (a) Total cross sections for charm production $\sigma_{c\bar 
c}$ up to NLO, for different PDFs, compared to the one used in the TIG model  
\cite{TIG} (for MRS R1 we also show the Born cross section). (b) Related $K_c 
$ factors. 
 
\item[Fig.~5]  $E^{3}$-weighted vertical prompt fluxes, for different PDFs, 
at NLO (for MRS R1 we also show the Born flux), for the three types of 
leptons considered, compared to the TIG \cite{TIG} conventional and prompt 
fluxes (left figures) and the related $K_l$ factors for each case (right 
figures). 
 
\item[Fig.~6]  (a) Contributions of the different Born and NLO processes to 
the total $E^{3}$-weighted vertical prompt fluxes. (b) Comparison of the 
fluxes with or without the showering option, at Born and NLO. (c) Comparison 
of the fluxes calculated in the `single' or `double' mode, at Born only 
(PDF: MRS R1). 
\end{description} 
 
\newpage 
 
\begin{table}[ptb] 
\begin{center} 
\begin{tabular}{|l|c|c|c|c|c|} 
\hline 
& Beam &  &  &  &  \\  
& Energy & $\sigma_{+}(x_{F} >0)$ & $\sigma_{c\bar c}$(EXP.) & $\sigma_{c 
\bar c}$(MNR) & PDF \\  
& (GeV) & $(\mu b)$ & $(\mu b)$ & $(\mu b)$ &  \\ \hline 
$pN$ & 250 & $\sigma_{+} (D^{+}, D^{-}) = 3.3 \pm0.4 \pm0.3$ & $13.5 \pm2.2$ 
& 13.54 & MRS R1 \\  
E769 \cite{alves1} &  & $\sigma_{+} (D^{0} ,\bar D^{0}) = 5.7 \pm1.3 \pm0.5$ 
&  & $m_{c}= 1.185$ GeV &  \\ \hline 
$pN$ & 250 & $\sigma_{+} (D^{+}, D^{-}) = 3.3 \pm0.4 \pm0.3$ & $13.5 \pm2.2$ 
& 13.43 & MRS R2 \\  
E769 \cite{alves1} &  & $\sigma_{+} (D^{0} ,\bar D^{0}) = 5.7 \pm1.3 \pm0.5$ 
&  & $m_{c}= 1.31$ GeV &  \\ \hline 
$pN$ & 250 & $\sigma_{+} (D^{+}, D^{-}) = 3.3 \pm0.4 \pm0.3$ & $13.5 \pm2.2$ 
& 13.59 & CTEQ4M \\  
E769 \cite{alves1} &  & $\sigma_{+} (D^{0} ,\bar D^{0}) = 5.7 \pm1.3 \pm0.5$ 
&  & $m_{c}= 1.27$ GeV &  \\ \hline 
$pN$ & 250 & $\sigma_{+} (D^{+}, D^{-}) = 3.3 \pm0.4 \pm0.3$ & $13.5 \pm2.2$ 
& 13.45 & CTEQ3M \\  
E769 \cite{alves1} &  & $\sigma_{+} (D^{0} ,\bar D^{0}) = 5.7 \pm1.3 \pm0.5$ 
&  & $m_{c}= 1.24$ GeV &  \\ \hline 
\end{tabular} 
\end{center} 
\caption{ Data on total cross sections for charm production for $pN$ 
collisions, from E769 experiment, have been converted to ${c\bar{c}}$ cross 
sections and compared to the predictions of the MNR program running at 
slightly different values of the charm mass m$_{c}$, using different PDFs.} 
\label{table:1} 
\end{table} 
 
\begin{table}[ptb] 
\begin{center} 
\begin{tabular}{|l|c|c|c|c|c|} 
\hline 
& Beam &  $\sigma_{+}(x_{F} >0)$ &  $\sigma_{c\bar c}$(EXP.) & $\sigma_{c \bar c}$(MNR) & $\sigma_{c \bar c}$(MNR) \\  
& Energy & $(\mu b)$ & $(\mu b)$ & 
$(\mu b)$ & $(\mu b)$ \\  
& (GeV) & & & $m_{c}=1.185$GeV &  $m_{c}=1.250$GeV  \\ \hline 
$\pi^{-} N$ & 210 & $D^{+} , D^{-}: 1.7 \pm0.3 \pm0.1$ & $9.7 
\pm1.2$ & 14.08 & 10.64 \\  
E769 \cite{alves1} &  & $D^{0} , \bar D^{0}: 6.4 \pm0.9 \pm0.3$ \hfil 
&  &  &  \\ \hline 
$\pi^{-} N$ & 250 & $D^{+} , D^{-}: 3.6 \pm0.2 \pm0.2$ & $14.2 
\pm1.1$  \hfil & 16.54 & 12.56 \\  
E769 \cite{alves1} &  & $D^{0} ,\bar D^{0}: 8.2 \pm0.7 \pm0.5$  \hfil 
&  &  &  \\ \hline 
$\pi^{+} N$ & 250 & $D^{+} , D^{-}: 2.6 \pm0.3 \pm0.2$ & $10.0 
\pm1.2$  \hfil & 16.54 & 12.56 \\  
E769 \cite{alves1} &  & $D^{0} , \bar D^{0}: 5.7 \pm0.8 \pm0.4$  \hfil 
&  &  &  \\ \hline 
$\pi^{\pm} N $ & 250 & $D^{+} , D^{-}: 3.2 \pm0.2 \pm0.2$ & $ 
12.5 \pm0.8$  \hfil & 16.54 & 12.56 \\  
E769 \cite{alves1} &  & $D^{0} , \bar D^{0}: 7.2 \pm0.5 \pm0.4$  \hfil 
&  &  &  \\ \hline 
$\pi^{-} N$ & 350 & $D^{+} , D^{-}: 3.28 \pm0.08 \pm0.29$ & $ 
13.3 \pm0.7$  \hfil & 22.22 & 17.06 \\  
WA92 \cite{ada} &  & $D^{0} ,\bar D^{0}: 7.78 \pm0.14 \pm0.52$  \hfil 
&  &  & (13.5 for $m_c$\\  
&  &  &  &  & =1.31GeV) \\ \hline 
\end{tabular} 
\end{center} 
\caption{Data on total cross sections for charm production for $\protect\pi N 
$ collisions, from E769 and WA92 experiments, have been converted to ${c\bar{ 
c}}$ cross sections and compared to the predictions of the MNR program 
running at slightly different values of the charm mass m$_{c}$, using MRS R1. 
} 
\label{table:2} 
\end{table} 

\newpage
\begin{figure}[t]
\epsfig{file=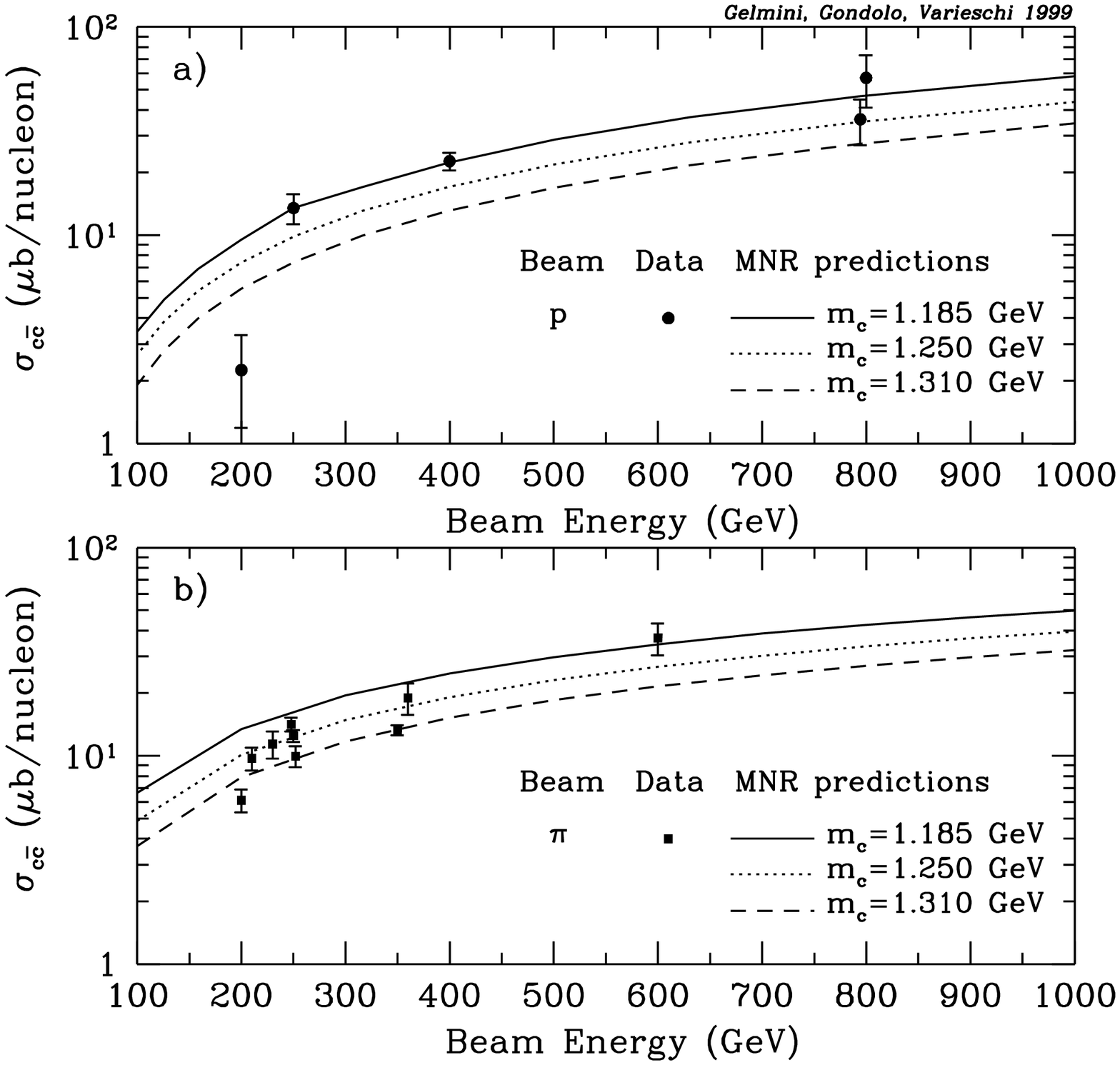,width=\textwidth}
\caption{~}
\end{figure}

\newpage
\begin{figure}[t]
\epsfig{file=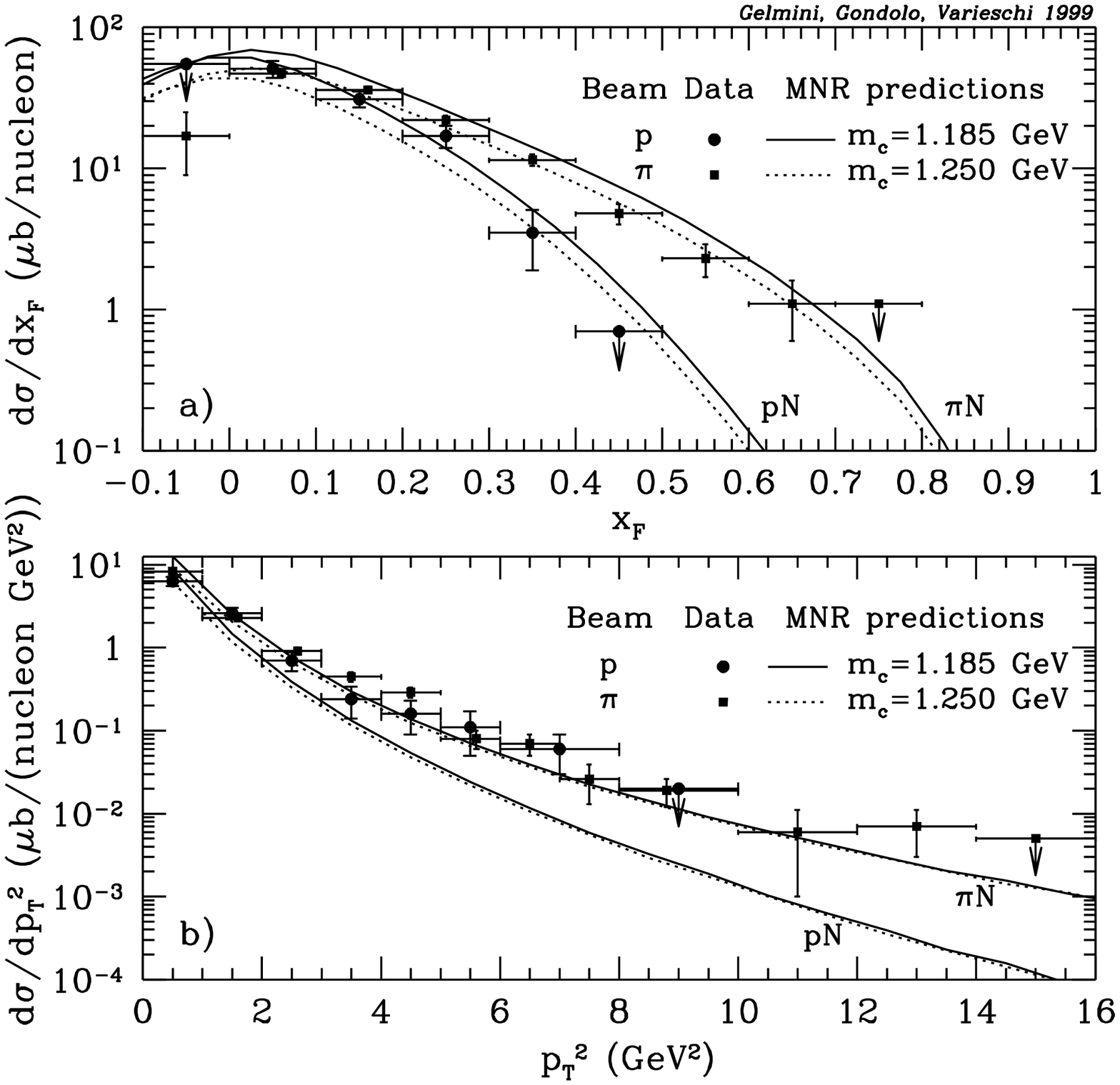,width=\textwidth}
\caption{~}
\end{figure}

\newpage
\begin{figure}[t]
\epsfig{file=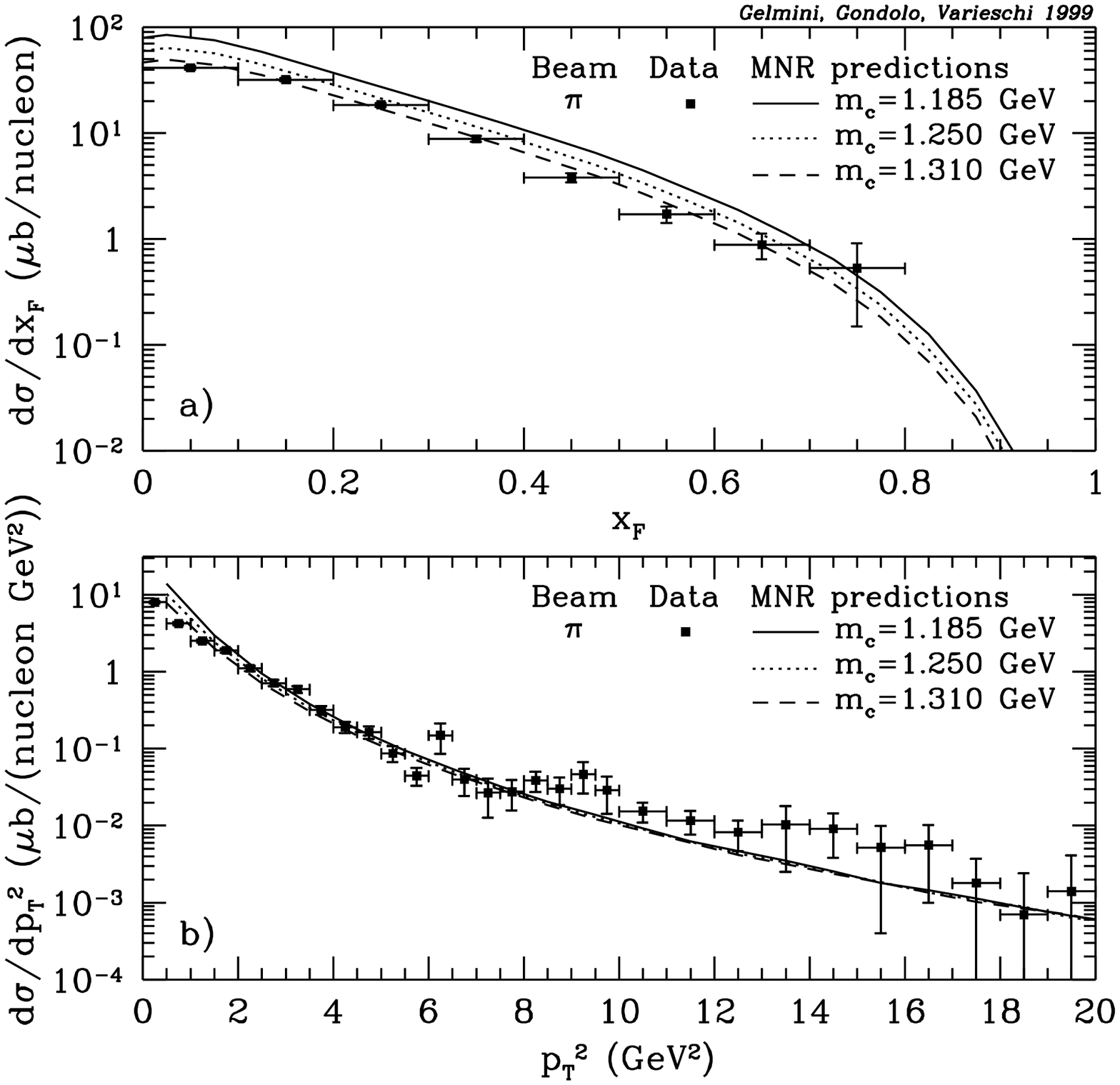,width=\textwidth}
\caption{~}
\end{figure}

\newpage
\begin{figure}[t]
\epsfig{file=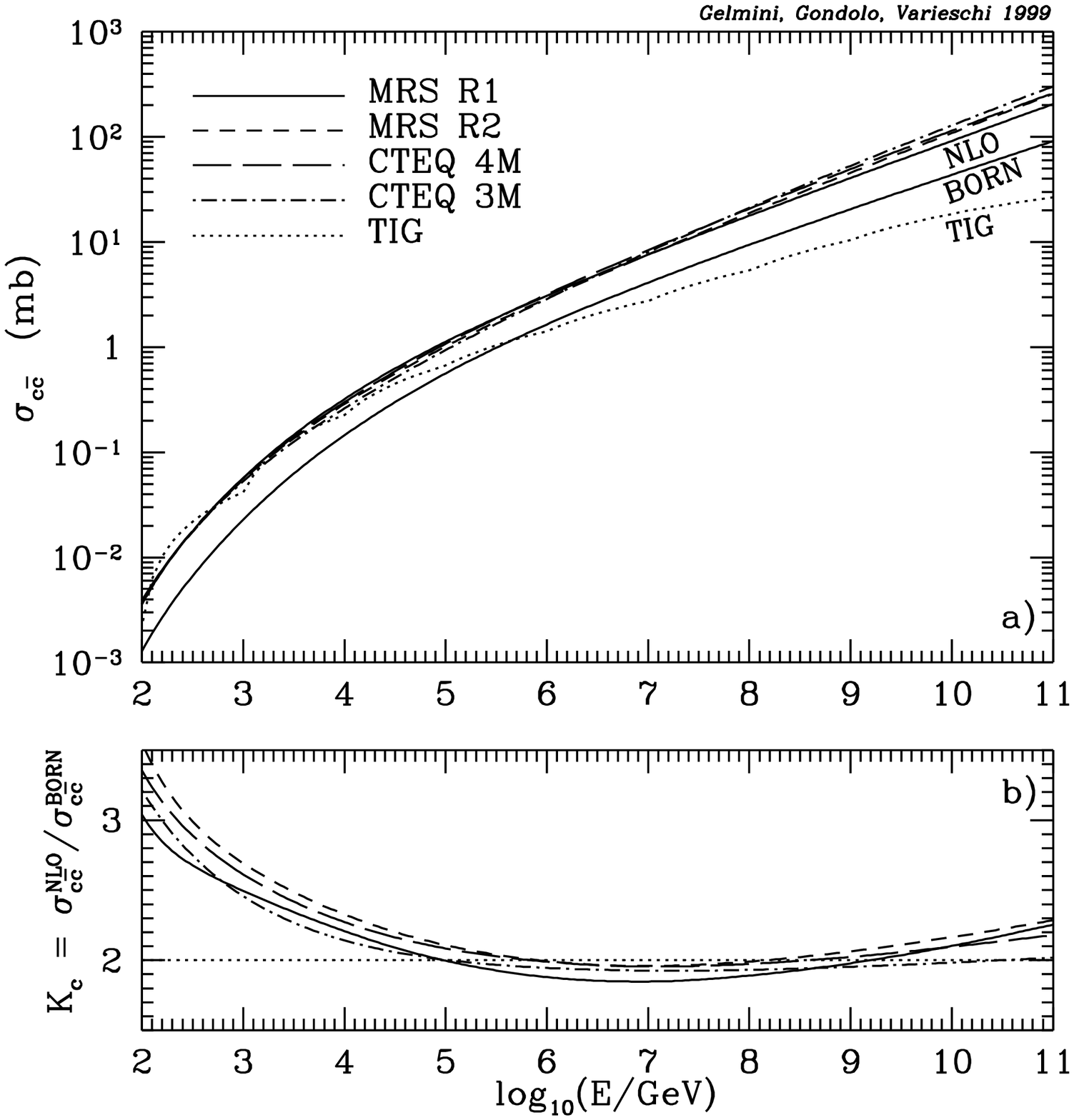,width=\textwidth}
\caption{~}
\end{figure}

\newpage
\begin{figure}[t]
\epsfig{file=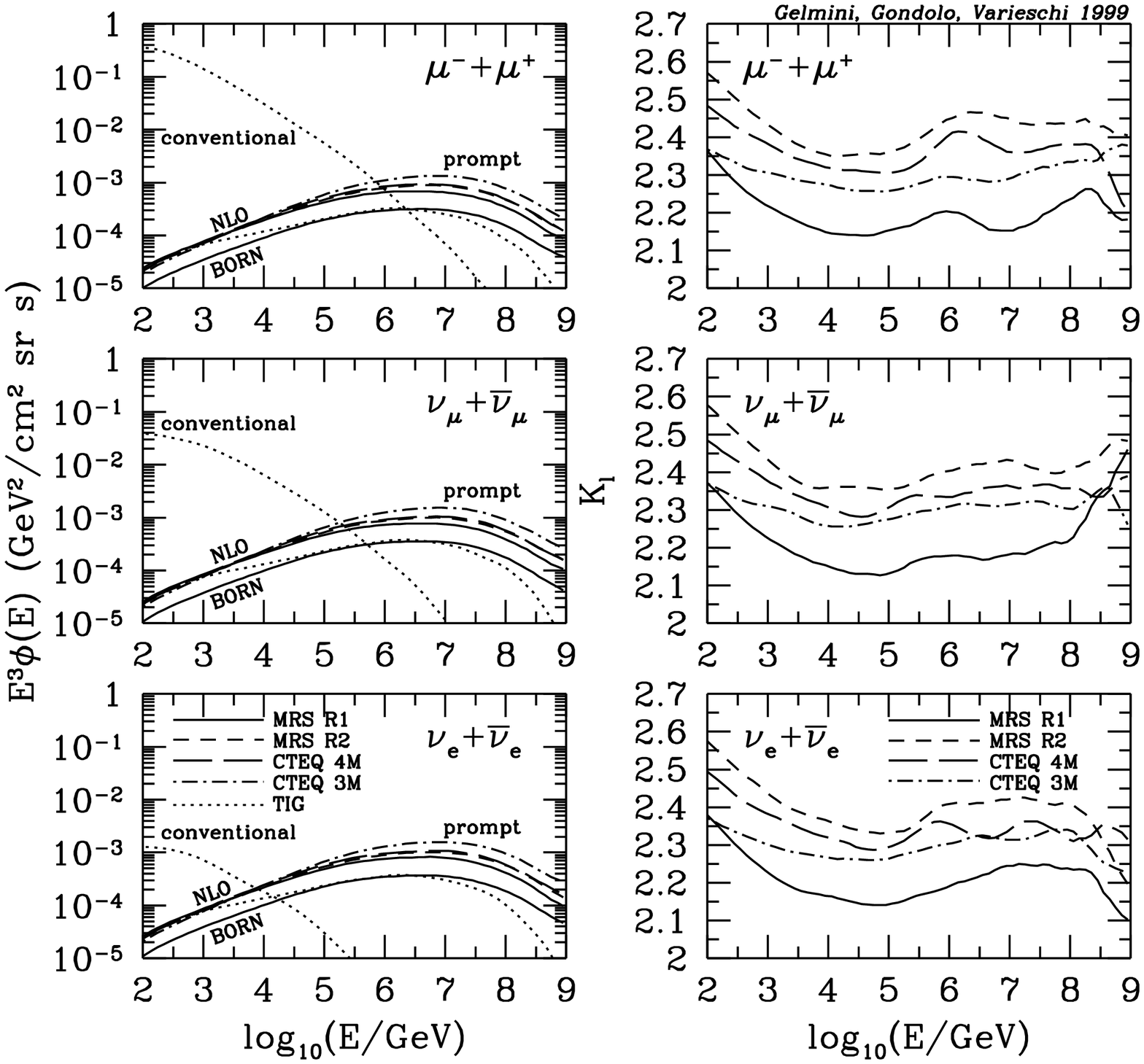,width=\textwidth}
\caption{~}
\end{figure}

\newpage
\begin{figure}[t]
\epsfig{file=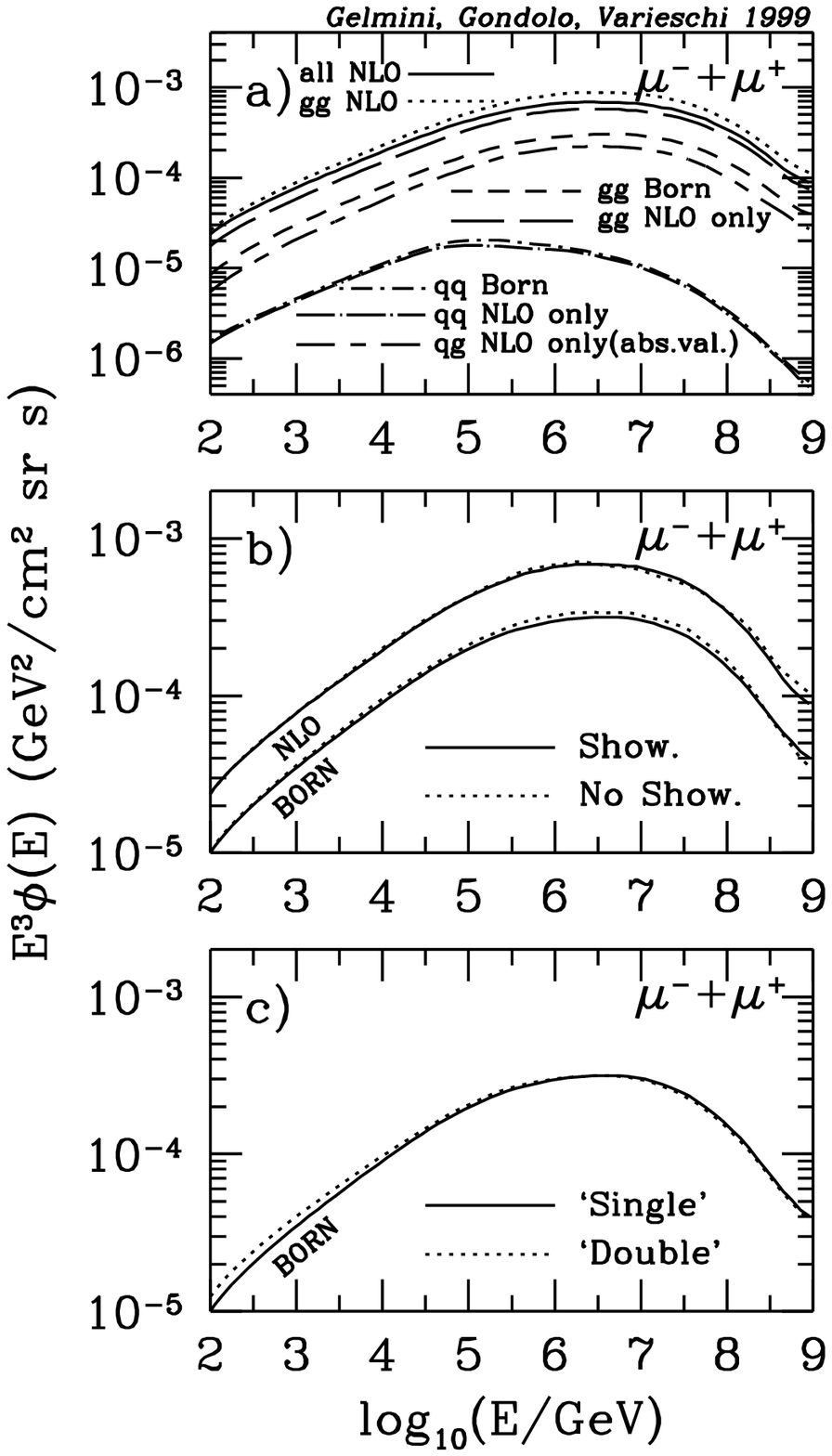,width=\textwidth}
\caption{~}
\end{figure}

\end{document}